\documentclass[aps,twocolumn,pra,superscriptaddress,showpacs,tightenlines]{revtex4-1}
\usepackage{amsmath}
\usepackage{graphicx}
\usepackage{epsfig}
\usepackage{amsfonts}
\usepackage{color}
\usepackage[colorlinks,citecolor=blue]{hyperref}

\begin{document}

\title{Enhanced interferometry using squeezed thermal states and even or odd states}
\author{Qing-Shou Tan}
\affiliation{CEMS, RIKEN, Saitama 351-0198, Japan}
\affiliation{Zhejiang Institute of Modern Physics, Department of Physics, Zhejiang
University, Hangzhou 310027, China}
\author{Jie-Qiao Liao}
\affiliation{CEMS, RIKEN, Saitama 351-0198, Japan}
\author{Xiaoguang Wang}
\affiliation{Zhejiang Institute of Modern Physics, Department of Physics, Zhejiang
University, Hangzhou 310027, China}
\author{Franco Nori}
\affiliation{CEMS, RIKEN, Saitama 351-0198, Japan}
\affiliation{Department of Physics, The University of Michigan, Ann Arbor, Michigan
48109-1040, USA}
\date{\today }

\begin{abstract}
We derive a general expression of the quantum Fisher information for a
Mach-Zehnder interferometer, with the port inputs of an \emph{arbitrary}
pure state and a squeezed thermal state. We find that the standard quantum
limit can be beaten, when even or odd states are applied to the pure-state
port. In particular, when the squeezed thermal state becomes a thermal
state, all the even or odd states have the same quantum Fisher information for given photon numbers.
For a squeezed thermal state, optimal even or odd
states are needed to approach the Heisenberg limit. As examples,
we consider several common even or odd states: Fock states, even or odd coherent states, squeezed
vacuum states, and single-photon-subtracted squeezed vacuum states.
We also demonstrate that super-precision can be realized by implementing the
parity measurement for these states.
\end{abstract}



\maketitle

\section{Introduction}

Interferometers are extremely useful and precise measuring tools,
which have been widely used to estimate very small phase changes in quantum
metrology~\cite{Caves1981,Yurke1985,Holland1993,Giovannetti2011,Dorner2009,Sanders1995,Ma2011,Ma2011B,Humphreys2013,Dowling2008,Lvovsky2009,Katori2011,Shevchenko2010}.
In general, the sensitivity of phase estimation within these settings
crucially depends on the input states as well as the detection schemes. For
a Mach-Zehnder interferometer (MZI) with classical-light inputs, the phase
sensitivity is bounded by the standard quantum limit (SQL) (also called
shot-noise limit), i.e., $1/\sqrt{n_{T}}$, where $n_{T}$ is the total photon number~
\cite{Helstrom1976,Holevo1982}. To go beyond the SQL in MZI, entangled
states (the states after the first beam splitter) are usually needed to
carry the phase information.

It has been shown that, with parity measurements~\cite
{Gerry2000,Gerry2001,Gerry2010,Gao2010,Anisimov2010,plick2010,Chiruvelli2011,Seshadreesan2011,Sahota2013,Zhang2013}, the Heisenberg limit (HL), i.e., $1/n_{T}$, can be achieved in lossless
interferometers by using maximally path-entangled states, such as NOON
states~\cite{Bollinger1996,Boto2000,Hofmann2007,Huver2008} and entangled
coherent states~\cite{Gerry2001,Joo2011,Jin2013} (here the first beam splitter of the MZI should be replaced
by devices to generate the entangled states). However, from the viewpoint of current experimental technology, it is a hard task to
produce these entangled states involving a large number of photons. Due to
restrictions in the photon numbers which can be reached, the estimation precision with
maximally entangled states is even possibly worse than that obtained with
high-intensity classical light sources~\cite{Escher2011,Lang2013}. Given
this situation, finding optimal high-intensity states as inputs of MZIs
is of practical relevance.

As an example of how to enhance the phase sensitivity with high-intensity
input states, Caves~\cite{Caves1981} considered the inputs of a
high-intensity coherent state and a low-intensity squeezed vacuum state.
Since then, many theoretical and experimental studies have focused on this
topic~\cite{Pezze2008,Ono2010,Seshadreesan2011,Jarzyna2012,Lang2013}. More
recently, an alternative method to reach a
sub-shot-noise phase uncertainty with high-intensity states has been studied in Ref.~\cite{Pezze2013}.
They~\cite{Pezze2013} considered the configuration where the MZI is fed by a Fock
state in one port and a high-intensity state, either a coherent state or a
thermal state, in the other port. We note that, in these two cases, the
separated-input states will be entangled to carry the phase by the first
beam splitter of the MZI.

Here, we consider a more general case, in which the input state of the MZI is
\begin{equation}
\rho_{\rm{in}}=|\psi\rangle_{a\;a}\langle\psi|\otimes\rho_{b},
\label{initialstate}
\end{equation}
where $|\psi\rangle_{a}$ is an \emph{arbitrary} pure state and $\rho_{b}$ is
a squeezed thermal state~\cite{Kim1989}
\begin{equation}
\rho_{b}=\sum_{n=0}^{\infty} \frac{\bar{n}^{n}_{\rm{th}}}{(\bar{n}_{\rm{
th}}+1)^{n+1}}S_{b}(\xi)|n\rangle _{b\,b}\langle n|S^{\dagger}_{b}(\xi),\label{squethermstat}
\end{equation}
with the average thermal photon number $\bar{n}_{\text{th}}$. The squeezing
operator is defined by $S_{b}(\xi)=\exp [(-\xi b^{\dagger 2}+\xi^{\ast }b^{2})/2]$, with
the squeezing factor $\xi=re^{i\theta}$ (hereafter we choose $\theta=0$).
The squeezed thermal state $\rho_{b}$ can be generated by either injecting a
thermal field to a squeezing device or passing a squeezed vacuum state
through a thermal noise channel. Mathematically, the scenario under
consideration covers several special cases of significance. (i) When $\bar{n}_{\rm{th}}=0$,
$\rho_{b}$ is a squeezed vacuum state. If we further choose
$|\psi \rangle_{a}$ as a coherent state, then the state $\rho_{\text{in}}$
is reduced to the input state in Ref.~\cite{Caves1981}. (ii) When $r=0$, the
squeezed thermal state is reduced to a thermal state; if we now choose $
|\psi \rangle_{a}$ as a Fock state, then this produces a special case of the
state discussed in Ref.~\cite{Pezze2013}.

We note that the quantum Fisher information (QFI), which is related to the quantum Cram\'{e}r-Rao bound (CRB)~\cite{Braunstein1996}, has been widely used in quantum metrology~\cite{Anisimov2010,Joo2011,Jin2013,Escher2011,Lang2013,Ono2010,Jarzyna2012,Pezze2008,Pezze2013}.
For example, the QFI has been used to characterize the phase sensitivity when the MZI is fed by
a two-mode squeezed vacuum state~\cite{Anisimov2010}, an entangled coherent state after the first beam splitter~\cite{Joo2011,Jin2013},
and a coherent state together with a squeezed vacuum state~\cite{Pezze2008,Lang2013}.
In this work, we will also describe the phase sensitivity with the QFI. By calculating the QFI, we find that, if $|\psi\rangle_{a}$ is composed of
either only-even or only-odd number states, the SQL for the phase-shift
measurement can be beaten even when there is no squeezing in the thermal
state. If there is squeezing in the thermal state, the HL can be approached
for certain even or odd states. As examples, we consider Fock states, even or odd coherent states, squeezed vacuum states, and
single-photon-subtracted squeezed vacuum states. Furthermore, we
consider the photon-number parity measurement~\cite{Gerry2000,Gerry2001,Gerry2010,Gao2010,Anisimov2010,plick2010,Chiruvelli2011,Seshadreesan2011,Sahota2013,Zhang2013},
which was introduced into optical interferometry in Refs.~\cite{Gerry2000,Gerry2001}.
Recently, it was shown in Ref.~\cite{Seshadreesan2011} that,
in two-path optical interferometry, the photon-number parity measurement achieves
the quantum CRB of phase sensitivity for all proposed pure states in the field of sub-shot-noise phase sensitivity.
In this work, our results indicate that, for the even and odd states considered here, the quantum CRB can also be reached by implementing the parity measurement.

\section{Quantum Fisher information in MZ interferometers}

\begin{figure}[tbp]
\includegraphics[bb=80 650 372 735, width=3. in]{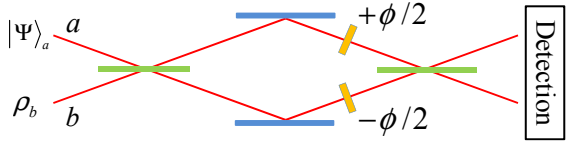}
\caption{(Color online) Schematic diagram of a balanced MZI, which is
composed of two $50$:$50$ beam splitters (shown in green) and two phase
shifters (orange). The input state of the two ports is $|\protect\psi
\rangle _{a\;a}\langle \protect\psi |\otimes \protect\rho _{b}$, where $|
\protect\psi \rangle _{a}$ is an arbitrary pure state and $\protect\rho _{b}$
is a squeezed thermal state.}
\label{setup}
\end{figure}
The balanced MZI considered here is formed by two $50$:$50$ beam splitters
and two phase shifters, as shown in Fig.~\ref{setup}. The two input ports
are fed by the state $\rho _{\text{in}}$ given in Eq.~(\ref{initialstate}).
If we denote the bosonic-mode annihilation operators of the two ports as $a$
and $b$, then the unitary transformation associated with this interferometer
can be written as
\begin{equation}
U(\phi)=e^{-i(\pi/2)J_{x}}e^{i\phi J_{z}}e^{i(\pi/2)
J_{x}}=\exp(-i\phi J_{y}),\label{unievolopMZI}
\end{equation}
where $\phi $ is the phase to be estimated. The operators
\begin{subequations}
\begin{align}
J_{x}&=\frac{1}{2}(a^{\dag}b+b^{\dag }a),\\
J_{y}&=-\frac{i}{2}(a^{\dag }b-b^{\dag }a),\\
J_{z}&=\frac{1}{2}(a^{\dag}a-b^{\dag }b)
\end{align}
\end{subequations}
are the usual angular momentum operators in the Schwinger
representation. These satisfy the commutation relations $
[J_{x},J_{y}]=iJ_{z} $, $[J_{y},J_{z}]=iJ_{x}$, and $[J_{z},J_{x}]=iJ_{y}$.

Before addressing the case of our input state $\rho_{\text{in}}$, we first
give the QFI for a general separable-state input: $\rho_{a}\otimes\rho_{b}$,
where $\rho_{a}$ and $\rho_{b}$ could be either pure states or mixed states.
For this input state, the output state is
$\rho_{\mathrm{out}}=U(\phi)\rho_{a}\otimes\rho_{b}U^{\dag}(\phi)$,
and the ultimate limit of
phase sensitivity is given by the quantum CRB~\cite{Braunstein1996},
\begin{equation}
\Delta\phi_{\text{min}}=1/\sqrt{\mathcal{F}},\hspace{1 cm} \mathcal{F}=\text{
Tr}(\rho_{\text{out}}G^{2}),  \label{QCRvsQFI}
\end{equation}
where $\mathcal{F}$ is the QFI, with $G$ the optimal phase estimator. The
symmetric logarithmic derivation $G$ of the density matrix $\rho_{\text{out}
} $ is defined by the operator relation
\begin{equation}
\frac{\partial\rho_{\text{out}}}{\partial\phi}=\frac{1}{2}(\rho_{\text{out}}G+G\rho_{\text{out}}).
\end{equation}
Utilizing the
spectral decompositions $\rho_{a}=\sum_{j}p_{j}|\psi_{j}\rangle_{a\;a}
\langle \psi_{j}|$ and $\rho_{b}=\sum_{m}q_{m}|\varphi_{m}\rangle_{b\;b}
\langle \varphi_{m}|$, the QFI can be obtained as~\cite
{Helstrom1976,Holevo1982, Knysh2011,Liu2013}
\begin{equation}
\mathcal{F}=\sum_{k}4Q_{k}\langle\phi_{k}\vert
J_{y}^{2}\vert\phi_{k}\rangle-\sum_{kk^{\prime}}\frac{8Q_{k}Q_{k^{\prime}}}{
Q_{k}+Q_{k^{\prime}}}\vert\langle\phi_{k}\vert
J_{y}\vert\phi_{k^{\prime}}\rangle\vert^{2},  \label{f1}
\end{equation}
where $Q_{k}=p_{j}q_{m}$, and $\{|\phi _{k}\rangle
=|\psi_{j}\rangle_{a}\otimes|\varphi_{m}\rangle_{b}\}$ is a complete-set
basis in the two-mode Hilbert space. We should point out that the $\mathcal{F
}$ does not depend on the parameter $\phi$. For the pure-state case, Eq.~(\ref{f1}) is
reduced to $\mathcal{F}=4(\langle J_{y}^{2}\rangle_{\textrm{in}}-\langle J_{y}\rangle^{2}_{\text{in}})$.

When the input states on ports $a$ and $b$ are an arbitrary pure state $
|\psi\rangle_{a}$ and a squeezed thermal state $\rho_{b}$, respectively, the
QFI can be obtained as
\begin{equation}
\mathcal{F}_{|\psi\rangle_{a}}=\bar{n}_{a}+\bar{n}_{b}+2\bar{n}_{a}\bar{n}
_{b}+\Theta_{|\psi\rangle_{a}},  \label{QFI}
\end{equation}
where $\bar{n}_{a}=\,_{a}\langle\psi|a^{\dag}a|\psi\rangle_{a}$ and
\begin{equation}
\bar{n}_{b}=(2\bar{n}_{\text{th}}+1)\sinh^{2}(r)+\bar{n}_{\textrm{th}},
\end{equation}
are the average
photon numbers for modes $a$ and $b$, respectively. The $\Theta_{|\psi
\rangle_{a}}$ in Eq.~(\ref{QFI}) is given by
\begin{eqnarray}
\Theta_{|\psi\rangle_{a}} &=&\sinh (2r)(2\bar{n}_{\text{th}}+1)\text{Re}
[\langle a^{2}\rangle]-\frac{4\bar{n}_{\text{th}}(\bar{n}_{\text{th}}+1)}{2
\bar{n}_{\text{th}}+1}  \notag \\
&&\times \lbrack \cosh(2r)+\cos (2\varphi_{0})\sinh (2r)]|\langle a\rangle
|^{2},  \label{Theta}
\end{eqnarray}
where $\varphi _{0}$ is defined by $\langle a\rangle =|\langle a\rangle
|e^{i\varphi_{0}}$, and the expectation values $\langle a\rangle$ and $
\langle a^{2}\rangle$ are taken over the state $|\psi\rangle_{a}$.
Equations~(\ref{QFI}) and (\ref{Theta}) show that the QFI depends not only
on the average photon numbers of the two modes, but also on the statistical
properties of the annihilation operator: $\langle a\rangle$ and $\langle
a^{2}\rangle$.

Based on Eqs.~(\ref{QCRvsQFI}) and~(\ref{QFI}), we can determine the QFIs
corresponding to the SQL and HL. For an ideal MZI, the total photon number
operator $a^{\dagger}a+b^{\dagger}b$ is a conserved quantity. If we denote
the total photon number as $n_{T}\equiv\bar{n}_{a}+ \bar{n}_{b}$, then the
SQL and HL are defined by
\begin{equation}
\Delta\phi_{\text{SQL}}=1/\sqrt{n_{T}},\hspace{0.5 cm}\Delta\phi_{
\text{HL}}=1/n_{T}.
\end{equation}
In these two limits, the corresponding QFIs are
\begin{equation}
\mathcal{F}_{\text{SQL}}=n_{T}, \hspace{1 cm}\mathcal{F}_{\text{HL}}=n_{T}^{2}.
\end{equation}
Comparing $\mathcal{F}_{\text{SQL}}$ and $\mathcal{F}_{\text{
HL }}$ with Eq.~(\ref{QFI}), we can obtain these results: To surpass the
SQL, the condition $\Theta_{|\psi\rangle_{a}}>-2\bar{n}_{a}\bar{n}_{b}$
needs to be satisfied; while to approach the HL, the input state should
impose that $\Theta_{|\psi\rangle_{a}}\rightarrow\bar{n}_{a}^{2}+\bar{n}
_{b}^{2}-(\bar{n}_{a}+\bar{n}_{b})$.

For a fixed $n_{T}$, we expect to obtain a large $\mathcal{F}
_{|\psi\rangle_{a}}$, namely a small $\Delta\phi_{\text{min}}$, by choosing
a proper state $|\psi\rangle_{a}$. When $\bar{n}_{a}$ and $\bar{n}_{b}$ are
fixed, this means that we need to find some input states to make $
\Theta_{|\psi\rangle_{a}}$ as large as possible. In general, it is difficult
to know how the value of $\Theta_{|\psi\rangle_{a}}$ depends on the
statistics of mode $a$. However, in the following special case, we can
obtain a nonnegative $\Theta_{|\psi\rangle_{a}}$: under the condition of
either $\langle a\rangle=0$ or $\bar{n}_{\mathrm{th}}=0$, the second term in
Eq.~(\ref{Theta}) disappears, and then we always have $\Theta_{|\psi
\rangle_{a}}\geq 0$. We can check that all even or odd states satisfy this
condition $\langle a\rangle=0$. This means that even or odd states can be used
as a resource to enhance the QFI.

\section{Phase sensitivity and parity measurement for even or odd states}

The state $\rho_{b}$ in Eq.~(\ref{squethermstat}) has two variables: the squeezing factor $r$ and
the thermal photon number $\bar{n}_{\text{th}}$. When $r=0$, the squeezed
thermal state is reduced to a thermal state. In this case, to surpass the
SQL, a quantum state on port $a$ is needed. Recall that for the quasiclassical
(coherent) state $|\psi \rangle _{a}=|\alpha _{0}\rangle _{a}$, we always
have $\Theta _{|\alpha _{0}\rangle _{a}}=-2\bar{n}_{a}\bar{n}_{b}(\bar{n}
_{b}+1)(\bar{n}_{b}+1/2)^{-1}\leq -2\bar{n}_{a}\bar{n}_{b}$. In particular,
when $|\psi \rangle _{a}$ is an even or odd state, regardless of its form, the
QFI is
\begin{equation}
\mathcal{F}_{\mathrm{e/o}}=\bar{n}_{a}+\bar{n}_{b}+2\bar{n}_{a}\bar{n}_{b},
\end{equation}
where $\bar{n}_{b}=\bar{n}_{\text{th}}$. This result means that we can
obtain a sub-shot-noise uncertainty just by mixing a thermal light with an
\emph{arbitrary} even or odd state in a MZI. In particular, when $\bar{n}_{a}\sim\bar{n}_{b}\sim n_{T}/2\gg1$,
we have the approximate relation $\mathcal{F}_{\mathrm{e/o}}\propto n_{T}^{2}/2$, which is of the
same scale of $\mathcal{F}_{\textrm{HL}}=n_{T}^{2}$.
The situation for $r>0$ is more
complicated. The QFI in this case depends on the form of $|\psi \rangle _{a}$.
Below we will consider several common even or odd states: the Fock state $|N\rangle
_{a}$, even or odd coherent states $|\alpha _{0\pm }\rangle_{a}$, the squeezed vacuum
state $|\xi_{0}\rangle_{a}$, and the single-photon-subtracted squeezed vacuum
state $|\zeta(1)\rangle_{a}$. To see the
advantages of even or odd states, we first consider the coherent
state $|\alpha_{0}\rangle_{a}$ as a reference.

\subsection{Coherent state $|\alpha _{0}\rangle _{a}$}

Suppose that the port $a$ is fed by a coherent state $|\alpha _{0}\rangle
_{a}$ with $\alpha _{0}=|\alpha _{0}|e^{i\theta _{c}}$. When $\theta _{c}=0$
the QFI can be obtained from Eqs.~(\ref{QFI}) and (\ref{Theta}) as
\begin{equation}
\mathcal{F}_{|\alpha _{0}\rangle _{a}}=\frac{e^{2r}}{2\bar{n}_{\textrm{th}}+1}
\bar{n}_{a}+\bar{n}_{b},  \label{QFIcohe}
\end{equation}
where $\bar{n}_{a}=\bar{n}_{|\alpha _{0}\rangle _{a}}=|\alpha _{0}|^{2}$ and
$\bar{n}_{b}=(2\bar{n}_{\text{th}}+1)\sinh^{2}(r)+\bar{n}_{\textrm{th}}$. We
note that Eq.~(\ref{QFIcohe}) has been used to analyze the effects of linear
photon losses with inputs of coherent states and squeezed vacuum states~\cite
{Caves1981,Pezze2008,Lang2013}, and the quantum CRB can be obtained by measuring a
symmetric logarithmic derivative~\cite{Ono2010}. When $0\leq \bar{n}_{\rm{th}}<(e^{2r}-1)/2$, we have $\mathcal{F}_{|\alpha _{0}\rangle _{a}}>\mathcal{
F}_{\text{SQL}}$, then the SQL is surpassed. By analyzing the function
\begin{equation}
\Theta _{|\alpha _{0}\rangle _{a}}=|\alpha _{0}|^{2}\left[ \frac{\sinh (2r)-4
\bar{n}_{\text{th}}(\bar{n}_{\text{th}}+1)\cosh (2r)}{2\bar{n}_{\text{th}}+1}
\right] ,  \label{Thetacoh}
\end{equation}
we find $\Theta _{|\alpha _{0}\rangle _{a}}>0$ under the condition $0\leq
\bar{n}_{\text{th}}<(\sqrt{1+\tanh (2r)}-1)/2$. This relation shows that, in
the small thermal photon regime, the input with coherent state and squeezed thermal
state can surpass the QFI $\mathcal{F}_{\mathrm{e/o}}$. However, a
disadvantage in this case is that we cannot increase the total photon number
by adding the thermal photon number.

\subsection{Fock state $|N\rangle_a$}

In the case of the Fock state $|N\rangle _{a}$, the average photon number in mode $a$ is $N$. In this case, we have
\begin{equation}
\Theta _{|N\rangle_{a}}=0,
\end{equation}
and the QFI is~\cite{Pezze2013}
\begin{equation}
\mathcal{F}_{|N\rangle_{a}}=N+\bar{n}_{b}+2N\bar{n}_{b},
\end{equation}
which is independent of the values of $r$, for a
fixed average photon number $\bar{n}_{b}$. This means that the Fock state input can
naturally surpass the SQL. In particular, we have $\mathcal{F}_{|N\rangle
_{a}}>\mathcal{F}_{|\alpha_{0} \rangle _{a}}$ when $\bar{n}_{\text{th}}>(\sqrt{
1+\tanh (2r)}-1)/2$. Therefore, for a sufficiently large
thermal photon number $\bar{n}_{\text{th}}$, the Fock state is better than
coherent states for the estimation of phase uncertainty in our case.

The quantum CRB $\Delta\phi_{\text{min}}$ can be reached in
this case by detecting the photon number parity on one of the output modes.
For mode $a$, the photon-number parity operator is
\begin{equation}
\Pi_{a}=(-1)^{a^{\dag }\!a}.
\end{equation}
We can obtain the expectation value
of the parity operator by calculating the Wigner function of the
output state~\cite{Seshadreesan2011}. For the input state $|N\rangle
_{a\,a}\langle N|\otimes \rho _{b}$, the Wigner function is
\begin{equation}
W_{\mathrm{in}}(\alpha ,\beta )=W_{|N\rangle _{a}}(\alpha)\;W_{\rho_{b}}(\beta).
\end{equation}
Here $W_{|N\rangle _{a}}(\alpha )$ and $W_{\rho _{b}}(\beta )$ are, respectively,
the Wigner functions for the Fock state and the squeezed thermal state ($\theta =0$)~
\cite{Barnettbook}:
\begin{subequations}
\label{a1b1}
\begin{align}
W_{\left\vert N\right\rangle _{a}}(\alpha )& =\frac{2}{\pi }(-1)^{N}\exp
(-2\left\vert \alpha \right\vert ^{2})L_{N}(4\left\vert \alpha \right\vert
^{2}), \\
W_{\rho _{b}}(\beta )=& \frac{2}{\pi (2\bar{n}_{\mathrm{th}}+1)}\exp \left[ -
\frac{2(e^{2r}\beta _{r}^{2}+e^{-2r}\beta _{i}^{2})}{2\bar{n}_{\mathrm{th}}+1
}\right] ,
\end{align}
\end{subequations}
where $L_{N}(x)$ is the Laguerre polynomial of the $N$th order, $\beta _{r}$ and $\beta _{i}$ are the
real and imaginary parts of $\beta $, respectively. By making the replacements
\begin{subequations}
\begin{eqnarray}
\alpha &\rightarrow &\tilde{\alpha}=\alpha \cos (\phi /2)+\beta \sin (\phi/2), \\
\beta &\rightarrow &\tilde{\beta}=-\alpha \sin (\phi /2)+\beta \cos (\phi/2),
\end{eqnarray}
\end{subequations}
in $W_{\mathrm{in}}(\alpha,\beta)$, we obtain the Wigner
function of the output state as (see the Appendix for details)
\begin{equation}
W_{\mathrm{out}}(\alpha,\beta)=W_{\mathrm{in}}(\tilde{\alpha},\tilde{\beta}).\label{maprelation}
\end{equation}

The expectation value of the parity operator can be written as
\begin{equation}
\label{paritym}
\langle\Pi_{a}\rangle_{\vert N\rangle_{a}}=\frac{\pi}{2}\int_{-\infty}^{\infty}\!\!W_{\mathrm{out}}(0,\beta)\;d^{2}\!\beta,
\end{equation}
where the Wigner function at the origin of the phase space for mode $a$ is found to be
\begin{eqnarray}
W_{\mathrm{out}}(0,\beta ) &=&\frac{4(-1)^{N}}{\pi ^{2}(2\bar{n}_{\rm{th}}+1)}\exp
\left( -A\beta _{r}^{2}-B\beta _{i}^{2}\right)  \notag \\
&&\times L_{N}\left(4\sin^{2}(\phi/2)\vert\beta\vert^{2}\right)
\end{eqnarray}
with
\begin{subequations}
\label{definiAB}
\begin{eqnarray}
A &=&2\left[ \frac{e^{2r}}{2\bar{n}_{\rm{th}}+1}\cos ^{2}\left( \phi /2\right) +\sin
^{2}\left( \phi /2\right) \right] , \\
B &=&2\left[ \frac{e^{-2r}}{2\bar{n}_{\rm{th}}+1}\cos ^{2}\left( \phi /2\right) +\sin
^{2}\left( \phi /2\right) \right] .
\end{eqnarray}
\end{subequations}
It is a difficult task to write out the explicit form of Eq.~({\ref{paritym}}) for general $N$. However, for small $N$, the explicit form is accessible. When
$N=0,1,2$, we have
\begin{subequations}
\begin{eqnarray}
\left\langle \Pi _{a}\right\rangle _{\left\vert 0\right\rangle _{a}}&=&\frac{2}{(2\bar{n}_{\rm{th}}+1)\sqrt{AB}},\\
\langle\Pi_{a}\rangle_{\vert 1\rangle_{a}}&=&\frac{2\left[ (A+B)(1-\cos \phi
)-AB\right] }{(2\bar{n}_{\rm{th}}+1)(AB)^{3/2}}, \\
\left\langle \Pi _{a}\right\rangle_{\left\vert 2\right\rangle _{a}}&=&\frac{2\Xi }{(2\bar{n}_{\rm{th}}+1)(AB)^{5/2}},
\end{eqnarray}
\end{subequations}
where
\begin{eqnarray}
\Xi &=&AB[AB-2(A+B)(1-\cos \phi )]\nonumber\\
&&+2\sin ^{4}\left( \phi /2\right)
(3A^{2}+2AB+3B^{2}).
\end{eqnarray}

Using the result of $\langle\Pi_{a}\rangle$, we can obtain the fluctuation of the parity operator as $\Delta\Pi_{a}=\sqrt{\langle\Pi_{a}^{2}\rangle-\langle\Pi_{a}\rangle^{2}}=\sqrt{1-\langle\Pi_{a}\rangle^{2}}$.
According to the error propagation formula
\begin{equation}
\label{err}
\Delta\phi_{\mathrm{min}}=\min \left[ \frac{\Delta\Pi_{a}}{|d\langle\Pi_{a}\rangle/d\phi |}\right] ,
\end{equation}
we can analytically show that the quantum CRB $\Delta\phi_{\mathrm{min}}=1/\sqrt{\mathcal{F}_{|N\rangle _{a}}}$ (for $N=0,1,2$) can be reached in the limit $\phi
\rightarrow 0$. This result indicates that the quantum CRB can be reached by implementing the parity measurement. We also numerically checked that the quantum CRB can be reached with parity measurement for larger values of $N$.

\subsection{Even or odd coherent states $|\alpha_{0\pm}\rangle_{a}$}
We now turn to the case of  even or odd coherent states (also called Schr\"{o}dinger's cat states). The definition of  even or odd coherent states is
\begin{equation}
|\alpha _{0\pm }\rangle _{a}=\mathcal{N}_{\pm }^{\alpha _{0}}(|\alpha
_{0}\rangle \pm |-\alpha _{0}\rangle ),  \label{evenoddcohe}
\end{equation}
where $\mathcal{N}_{\pm }^{\alpha _{0}}=1/[2(1\pm
e^{-2|\alpha _{0}|^{2}})]^{1/2}$ are the normalization constants. Without loss of generality, hereafter we assume that $\alpha _{0}$ is real. According to Eq.~(\ref{Theta}), we
obtain
\begin{equation}
\Theta_{|\alpha _{0\pm }\rangle_{a}}=\alpha_{0}^{2}(2\bar{n}_{\text{th}
}+1)\sinh (2r).  \label{Thetaeven}
\end{equation}
The corresponding QFIs are then given by \label{QFIevenoddcohe}
\begin{equation}
\mathcal{F}_{|\alpha _{0\pm }\rangle _{a}}=(2\bar{n}_{\rm{th}}+1)[\alpha_{0}^{2}\sinh
(2r)+\bar{n}_{|\alpha_{0\pm }\rangle _{a}}\cosh (2r)]+\bar{n}_{b},
\end{equation}
where the average photon numbers are
\begin{subequations}
\begin{eqnarray}
\bar{n}_{|\alpha_{0+}\rangle_{a}} &=&\alpha_{0}^{2}\tanh (\alpha_{0}^{2}),\\
\bar{n}_{|\alpha_{0-}\rangle_{a}}&=&\alpha_{0}^{2}\coth (\alpha_{0}^{2}).
\end{eqnarray}
\end{subequations}
When $\alpha_{0}\geq2$, we have the approximate relation $\bar{n}_{|\alpha _{0+}\rangle _{a}}\simeq\bar{n}_{|\alpha_{0-}\rangle_{a}}\simeq \alpha_{0}^{2}$, and then the QFIs are approximately reduced to
\begin{equation}
\mathcal{F}_{|\alpha _{0+}\rangle _{a}}^{\prime }\simeq \mathcal{F}_{|\alpha
_{0-}\rangle _{a}}^{\prime }\simeq e^{2r}(2\bar{n}_{\text{th}}+1)\bar{n}_{|\alpha _{0\pm}\rangle _{a}}+
\bar{n}_{b},\label{approxi}
\end{equation}
We see from $\mathcal{F}_{|\alpha_{0\pm }\rangle _{a}}^{\prime }$ that a
sub-shot-noise uncertainty can be obtained as long as $r$ and $\bar{n}_{
\mathrm{th}}$ are not simultaneously zero. Furthermore, the HL can be
approached, if $\bar{n}_{\mathrm{th}}$ satisfies the relation
\begin{equation}
\bar{n}_{a}(2\bar{n}_{\mathrm{th}}+1)\sinh(2r)=\bar{n}_{a}^{2}+\bar{n}
_{b}^{2}-(\bar{n}_{a}+\bar{n}_{b}).
\end{equation}
According to $\mathcal{F}_{|\alpha_{0\pm }\rangle _{a}}^{\prime }$, we can
obtain a large QFI by increasing the average thermal photon number $\bar{n}_{
\mathrm{th}}$ for a fixed squeezing parameter $r$. This point is different
from the coherent-state case [Eq.~(\ref{QFIcohe})], in which a large $\bar{n}
_{\mathrm{th}}$ may lead to $\mathcal{F}_{|\alpha_{0}\rangle_{a}}<\mathcal{F}_{
\text{SQL}}$. It should be pointed out that these states are difficult
to be created with high photon numbers under current experimental conditions, but this might be possible in the future.
The Wigner function of the even or odd coherent state is~\cite{Gerrybook}
\begin{eqnarray}
W_{\vert\alpha_{0\pm}\rangle_{a}}(\alpha)&=&\frac{e^{-2|\alpha|^2}}{\pi \left(
1\pm e^{-2\alpha_{0}^{2}}\right) }\left[
e^{-2\alpha_{0}^{2}+4\alpha_{r}\alpha_{0}}\right.\nonumber\\
&&\left.+e^{-2\alpha_{0}^{2}-4\alpha_{r}\alpha_{0}}
\pm2\cos(4\alpha_{i}\alpha_{0})\right],\label{Wigfeocs}
\end{eqnarray}
where $\alpha_{r}$ and $\alpha_{i}$ are the real and imaginary parts of $\alpha$, respectively.
In terms of Eq.~(\ref{Wigfeocs}), we can obtain the expectation value of the parity operator, with the same method in the above section, as
\begin{eqnarray}
\langle \Pi_{a}\rangle_{|\alpha_{0\pm }\rangle_{a}} &=&\frac{2\left[e^{-2\alpha_{0}^{2}}\exp\left(\frac{
C^{2}}{4A}\right) \pm \exp\left(-\frac{C^{2}}{4B}\right)\right]
}{(2\bar{n}_{\rm{th}}+1)\left( 1\pm e^{-2\alpha_{0}^{2}}\right) \sqrt{AB}},\label{wxpPieocs}
\end{eqnarray}
where $A$ and $B$ have been given in Eq.~(\ref{definiAB}), and
\begin{equation}
C=4\alpha_{0}\sin \left( \phi /2\right) .
\end{equation}
Using Eqs.~(\ref{err}) and (\ref{wxpPieocs}), we can check that the quantum CRB
can be achieved when $\phi\rightarrow 0$, i.e., $\Delta \phi_{\mathrm{min}}=1/\sqrt{\mathcal{F}
_{|\alpha_{0\pm }\rangle_{a}}}$.
Similar analysis can be done for finite-dimensional even or odd coherent states~\cite{Miranowicz2014}.
\subsection{Squeezed vacuum state $|\xi_{0}\rangle_{a}$}
When the input state on port $a$ is a squeezed vacuum state $|\xi
_{0}\rangle _{a}=S_{a}(\xi _{0})|0\rangle _{a}$, with $\xi _{0}=Re^{i\theta
_{0}}$, which is an even state.  The average photon number for this state is $\bar{n}_{a}=\bar{n}_{|\xi _{0}\rangle_{a}}=\sinh^{2}(R)$. Based on the optimal phase-matching
condition $\theta _{0}=\pi $~\cite{Liu2013}, we obtain
\begin{equation}
\Theta _{|\xi _{0}\rangle _{a}}=(\bar{n}_{\mathrm{th}}+1/2)\sinh
(2R)\sinh (2r).
\end{equation}
The QFI in this case is
\begin{equation}
\mathcal{F}_{|\xi _{0}\rangle _{a}}=(\bar{n}_{\mathrm{th}}+1/2)\cosh
[2(R+r)]-1/2.  \label{QFIsqva}
\end{equation}
When $\bar{n}_{a}\gg 1$, Eq.~(\ref{QFIsqva}) is approximately reduced to
\begin{equation}
\mathcal{F}_{|\xi _{0}\rangle _{a}}^{\prime }\simeq e^{2r}(2\bar{n}_{
\mathrm{th}}+1)\bar{n}_{|\xi _{0}\rangle _{a}}+\bar{n}_{b},  \label{approxisvs}
\end{equation}
which has a similar form as in Eq.~(\ref{approxi}).
\begin{table*}[thb]
\begin{center}
\caption{The average photon number $\bar{n}_a$ in mode $a$, the function $\Theta_{|\psi\rangle_{a}}$, and the QFI $\mathcal{F}_{|\psi\rangle_{a}}$ for the MZI, when the two input ports are fed by an even or odd state $|\psi\rangle_{a}$ in mode $a$ and a squeezed thermal state $\rho_b$ in mode $b$.
Here, the $|\psi\rangle_{a}$ could be either a Fock state $|N\rangle_a$, even or odd coherent states $|\alpha_{0\pm}\rangle_{a}$ (here we assume $\alpha_0$ is real), squeezed vacuum state $|\xi_{0}\rangle_{a}$ with $\xi_{0}=-R$, or a single-photon-subtracted squeezed vacuum state $|\zeta(1)\rangle_{a}$ with $\zeta=-R'$. The average photon number in mode $b$ is $\bar{n}_{b}=(2\bar{n}_{\text{th}}+1)\sinh^{2}(r)+\bar{n}_{\textrm{th}}$, and the total photon number is $n_T=\bar{n}_a+\bar{n}_b$. The phase uncertainties can be obtained by $\Delta\phi_{\rm{min}}=1/\sqrt{\mathcal{F}_{|\psi\rangle_{a}}}$. For the states considered here, the SQL can be surpassed because of $\mathcal{F}_{|\psi\rangle_{a}}>n_{T}$, and the phase sensitivity $\Delta\phi_{\rm{min}}$ can be reached with the parity measurement.}
\begin{tabular}{|c|c|c|c|}
\hline
\hline
Input states $|\psi\rangle_{a}$ &$\bar{n}_a$& $\Theta_{|\psi\rangle_{a}}$& $\mathcal{F}_{|\psi\rangle_{a}}$  \\
\hline
$|N\rangle_a$ & $N$ &0& $N+(2N+1)[(2\bar{n}_{\text{th}}+1)\sinh^{2}(r)+\bar{n}_{\textrm{th}}]\;>\;n_T$  \\
\hline
$|\alpha_{0\pm}\rangle_{a}$&$\alpha_{0}^{2}$ for $\alpha_{0}\;\geq\;2$ &$\alpha_{0}^{2}(2\bar{n}_{\rm{th}}+1)\sinh (2r)$ & $\alpha_{0}^{2}(2\bar{n}_{\rm{th}}+1)\sinh (2r)\;>\;n_T$ \\
\hline
$|\xi_{0}\rangle_{a}$\; with\; $\xi_{0}=-R$ &$\sinh^2(R)$& $(\bar{n}_{\mathrm{th}}+1/2)\sinh
(2R)\sinh (2r)$ & $(\bar{n}_{\mathrm{th}}+1/2)\cosh
[2(R+r)]-1/2\;>\;n_T$ \\
\hline
$|\zeta(1)\rangle_{a}$\; with\; $\zeta=-R'$ &$1+3\sinh^2(R')$ & $3(\bar{n}_{\mathrm{th}}+1/2)\sinh (2R^{\prime
})\sinh (2r)$ & $3(\bar{n}_{\mathrm{th}}+1/2)\cosh
[2(R'+r)]-1/2\;>\;n_T$ \\
\hline
\hline
\end{tabular}
\end{center}\label{table11}
\end{table*}

Using the Wigner function of the squeezed vacuum state~\cite{Barnettbook}
\begin{equation}
W_{|\xi_{0}\rangle_{a}}(\alpha)=\frac{2}{\pi}\exp[-2(e^{-2R}\alpha
_{r}^{2}+e^{2R}\alpha_{i}^{2})],\label{Wigfsvs}
\end{equation}
the expected signal of the parity measurement can be obtained as
\begin{equation}
\langle \Pi_{a}\rangle_{|\xi_{0}\rangle_{a}}=\frac{2}{(2\bar{n}_{\rm{th}}+1)\sqrt{A_{1}B_{1}}},\label{expPisvs}
\end{equation}
where we introduce
\begin{subequations}
\label{a1b1}
\begin{align}
A_{1}=& \frac{2e^{2r}}{2\bar{n}_{\rm{th}}+1}\cos ^{2}(\phi
/2)+2e^{-2R}\sin ^{2}(\phi /2), \\
B_{1}=& \frac{2e^{-2r}}{2\bar{n}_{\rm{th}}+1}\cos ^{2}(\phi
/2)+2e^{2R}\sin ^{2}(\phi /2).
\end{align}
\end{subequations}
Based on Eqs.~(\ref{err}) and (\ref{expPisvs}), we obtain the quantum CRB for phase estimation
$\Delta\phi_{\rm{min}}=1/\sqrt{
\mathcal{F}_{|\xi_{0}\rangle _{a}}}$ in the limit $\phi\rightarrow 0$.

\begin{figure}[htbp]
\includegraphics[bb=5 29 249 393, width=3.4 in]{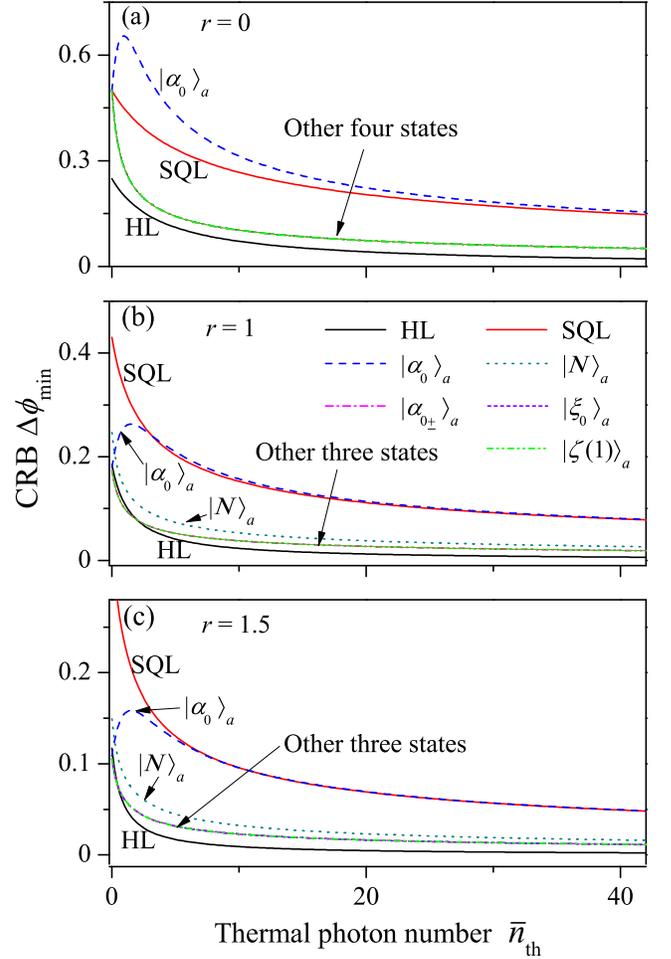}
\caption{(Color online) The quantum CRB $\Delta\phi_{\textrm{min}}\equiv(
\mathcal{F}_{|\psi\rangle_{a}})^{-1/2}$ as a function of the thermal photon number $\bar{n}_{
\mathrm{th}}$ for various input states in mode $a$: coherent state $|
\alpha_{0}\rangle_{a}$ (blue dashed curve), Fock state $|N\rangle_{a}$ (cadet blue dotted curve), even or odd coherent states $|\alpha_{0\pm}\rangle_{a}$ (magenta dash-dotted curve),
squeezed vacuum state $|\xi_{0}\rangle_{a}$ (purple short-dashed curve), and single-photon-subtracted
squeezed vacuum state $\vert\zeta(1)\rangle_{a}$ (green dash-dot-dotted
curve). Here, the average photon number for mode $a$ is $\bar{n}
_{a}\simeq4$, the mode $b$ is in the squeezed thermal state $\rho
_{b}$ with the squeezing factor (a) $r=0$, (b) $r=1$, (c) $r=1.5$. The HL (black solid
curve) and SQL (red solid curve) are shown for comparison. Note that the curves for the last four states exactly overlap with each
other in panel (a), and the
curves for the last three states approximately overlap with each
other in panels (b) and (c).}
\label{su2f12}
\end{figure}
\subsection{Single-photon-subtracted squeezed vacuum state $|\zeta(1)\rangle_{a}$}

The single-photon-subtracted squeezed vacuum state is defined by
\begin{eqnarray}
|\zeta (1)\rangle _{a}=\mathcal{N}_{1}aS_{a}(\zeta )|0\rangle _{a},
\end{eqnarray}
where $\zeta=R^{\prime}e^{i\theta^{\prime}}$, and $\mathcal{N}_{1}=1/\sinh(R^{\prime })$
is the normalization constant. Up to a trivial phase factor,
the state $|\zeta (1)\rangle _{a}$ is equivalent to the squeezed
single-photon state $S_{a}(\zeta )|1\rangle _{a}$, which has almost unit
fidelity to a superposed coherent state of small amplitude~\cite{jeong2005}.
We note that a state-input with squeezed single-photon state and coherent state has been studied in Ref.~\cite{Birrittella2012}.
When $\theta^{\prime}=\pi$, we have
\begin{equation}
\Theta_{|\zeta (1)\rangle_{a}}=3(\bar{n}_{\mathrm{th}}+1/2)\sinh (2R^{\prime
})\sinh (2r),
\end{equation}
and the average photon number $\bar{n}_{a}=\bar{n}_{|\zeta (1)\rangle _{a}}=1+3\sinh
^{2}(R^{\prime})$. According to Eq.~(\ref{QFI}), the QFI can be obtained as
\begin{equation}
\mathcal{F}_{|\zeta(1)\rangle_{a}}=3(\bar{n}_{\mathrm{th}}+1/2)\cosh[2(R^{\prime }+r)]-1/2,\label{QFI2}
\end{equation}
and the quantum CRB is $\Delta\phi_{\text{min}}=1/\sqrt{\mathcal{F}_{|\zeta(1)\rangle_{a}}}$.
When implementing the parity detection, we obtain
\begin{equation}
\langle\Pi_{a}\rangle_{|\zeta(1)\rangle_{a}}=\frac{2}{(2\bar{n}_{\mathrm{th}}+1)
\sqrt{A_{1}B_{1}}}\left(\frac{A_{2}}{2A_{1}}+\frac{
B_{2}}{2B_{1}}-1\right),\label{expPispssvs}
\end{equation}
where $A_{1}$ and $B_{1}$ are given by Eq.~(\ref{a1b1}) with the replacement
$R\rightarrow R^{\prime}$, $A_{2}$ and $B_{2}$ are defined by
\begin{equation}
A_{2}=4e^{-2R^{\prime}}\sin^{2}(\phi/2),\hspace{0.5cm}B_{2}=4e^{2R^{\prime}}\sin^{2}(\phi/2).
\end{equation}
In the derivation of Eq.~(\ref{expPispssvs}), we have used the Wigner function of the single-photon-subtracted squeezed vacuum state~\cite{Kenfack2004}
\begin{eqnarray}
W_{\vert \zeta(1)\rangle_{a}}(\alpha)&=&\frac{2}{\pi}
\exp[-2(e^{-2R'}\alpha^{2}_{r}+e^{2R'}\alpha^{2}_{i})]\nonumber\\
&&\times[4(e^{-2R'}\alpha^{2}_{r}+e^{2R'}\alpha^{2}_{i})-1].\label{Wignerspssvs}
\end{eqnarray}
In terms of Eqs.~(\ref{err}) and (\ref{expPispssvs}), we find that the best phase sensitivity,
i.e., the quantum CRB $\Delta \phi _{\mathrm{min}}=1/\sqrt{\mathcal{F}_{|\zeta
(1)\rangle_{a}}}$, can be reached in the limit $\phi\rightarrow 0$.

\subsection{Quantum CRB versus the thermal photon number $\bar{n}_{\rm{th}}$}

We now have obtained the phase sensitivity for a MZI, which is fed by various even or odd states in mode $a$ and a squeezed thermal state in mode $b$. To better show these results. Table I lists the average photon number, the function $\Theta_{|\psi\rangle_{a}}$, and the quantum Fisher information $\mathcal{F}_{|\psi\rangle_{a}}$ for various even or odd states: Fock states, even or odd coherent states, squeezed vacuum states, and single-photon-subtracted squeezed vacuum states. The phase uncertainties can be obtained by $\Delta\phi_{\rm{min}}=1/\sqrt{\mathcal{F}_{|\psi\rangle_{a}}}$. For the states considered here, the SQL can be surpassed because of $\mathcal{F}_{|\psi\rangle_{a}}>n_{T}$, and the phase sensitivity $\Delta\phi_{\rm{min}}$ can be reached with the parity measurement.

To clearly see the behaviors of the phase sensitivity for various input
states, in Fig.~\ref{su2f12} we plot the quantum CRB $\Delta \phi _{\text{min}}=1/
\sqrt{\mathcal{F}_{|\psi\rangle_{a}}}$ as a function of $\bar{n}_{\mathrm{th}}$. Here we fix
the average photon number in mode $a$: $\bar{n}_{a}\simeq 4$. Even or odd states used here with this photon number are accessible with current or near-future experiments.
For the Fock state $|N\rangle_{a}$, $\bar{n}_{a}\simeq 4$ means $N=4$. In experiments, Fock states with
up to two~\cite{Ourjoumtsev2006} and three~\cite{Cooper2013} running photons in optics have recently been generated.
For even or odd coherent states $|\alpha_{0\pm}\rangle_{a}$, $\bar{n}_{a}\simeq 4$ corresponds to $\alpha_{0}\simeq2$, which is very close to current experimental realizations. A recent experiment~\cite{Ourjoumtsev2007} reported the generation of an optical coherent-state superposition with $\alpha_{0}\simeq\sqrt{2.6}$.
For squeezed vacuum and single-photon states, we have $R\simeq1.45$ and $R^{\prime }\simeq 0.94$, satisfying the relation $\sinh^{2}(R)=1+3\sinh^{2}(R^{\prime})$ or $\cosh(2R)=3\cosh(2R^{\prime})$.
The squeezing factor $R=1.45$ corresponds to a squeezing of $12.6$ dB,
which can be realized with current experimental techniques~\cite{Vahlbruch2008}.

As shown in Fig.~\ref{su2f12}, when increasing $\bar{n}_{\mathrm{th}}$, the
phase uncertainties for even- or odd-state inputs decrease monotonically. In some parameters, the phase uncertainties can approach the HL. However, for the coherent-state case, the phase
uncertainty first increases and then gradually decreases, and it ultimately
approaches the SQL when increasing $\bar{n}_{\mathrm{th}}$.
When $r=0$ ($\rho _{b}$ becomes the thermal state),
we can see from Fig.~\ref{su2f12}(a) that the values of the optimal phase estimation (corresponding to
the quantum CRB) satisfy $\Delta\phi_{|\alpha_{0\pm}\rangle_{a}}=\Delta\phi_{|\xi_{0}\rangle_{a}}=\Delta\phi
_{|\zeta(1)\rangle_{a}}\leq\Delta\phi_{|\alpha\rangle_{a}}$, where
the last equality takes place if and only if $\bar{n}_{\mathrm{th}}=0$
(i.e., $\rho _{b}$ is a vacuum state). When $\rho_{b}$ is a thermal state,
the values of the optimal phase estimation for all even or odd states are the
same, which can beat the SQL. This point can be seen
from the expression of $\mathcal{F}_\mathrm{e/o}$. However, the phase
sensitivity for a coherent-state input cannot beat the SQL, because two
classical states cannot be entangled by a beam splitter~\cite{kim}.

When $r>0$, as shown in Figs.~\ref{su2f12}(b) and~\ref{su2f12}(c), the phase sensitivities for all
even or odd states always beat the SQL, and approach the HL when $\bar{n}_{b}\sim\bar{n}_{a}$.
In the coherent state case, the phase sensitivity
can beat the SQL when $\bar{n}_{\mathrm{th}}$ is not too large. This is
because $\mathcal{F}_{|\alpha _{0}\rangle _{a}}>\mathcal{F}_{\text{SQL}}$
when $0\leq \bar{n}_{\text{th}}<(e^{2r}-1)/2$. In addition, for a given state in  Fig.~\ref{su2f12} panels, from top to bottom, the values of the corresponding points
with the same $\bar{n}_{\mathrm{th}}$ decrease because the squeezing
part will increase the total photon number.

\section{Conclusion}

In summary, we have studied the QFI for a MZI, which is fed by an arbitrary
pure state and a squeezed thermal state. We have shown that, when the input
pure state is an even or odd state, the phase sensitivity can be drastically
improved. By mixing the even or odd states and a high-intensity thermal light,
a sub-shot-noise phase uncertainty can be obtained, and this uncertainty
only depends on the total photon number, regardless of the form of the
even or odd states. For the case of a squeezed thermal state, the sensitivity
can be further improved even approaching the HL when the pure-state port is
fed by an even or odd state. As examples, we considered Fock states,
even or odd coherent states, squeezed vacuum states, and single-photon-subtracted squeezed
vacuum states. Furthermore, we have demonstrated that the super-precision
given by the quantum CRB can be realized by implementing the parity measurement.

\begin{acknowledgments}
We would like to thank the referee for helpful suggestions. Q.S.T. is supported
by the China Postdoctoral Science Foundation (Grant No. 2013M541766). J.Q.L. is
supported by the JSPS Foreign Postdoctoral Fellowship No. P12503. X.W.
acknowledges support from the NFRPC through Grant No. 2012CB921602 and the
NSFC through Grants No. 11025527 and No. 10935010. F.N. is partially supported
by the RIKEN iTHES Project, MURI Center for Dynamic Magneto-Optics,
JSPS-RFBR Contract No. 12-02-92100, Grant-in-Aid for Scientific Research
(S), MEXT Kakenhi on Quantum Cybernetics, and the JSPS via its FIRST program.
\end{acknowledgments}

\appendix*
\section{Derivation of Eq.~(\ref{maprelation})}

In this Appendix, we present a detailed derivation of Eq.~(\ref{maprelation}).
The Wigner function of the input state $\rho_{\rm{in}}$ is defined by
\begin{eqnarray}
&&W_{\rm{in}}(\alpha,\beta)\nonumber\\
&=&\frac{4}{\pi^{2}}\textrm{Tr}[\rho_{\rm{in}}
D_{b}(\beta)D_{a}(\alpha)(-1)^{a^{\dagger}a+b^{\dagger}b}D_{a}^{\dagger}(\alpha)
D_{b}^{\dagger}(\beta)],\nonumber\\
\end{eqnarray}
where the displacement operators are
defined by $D_{a}(\alpha)=e^{\alpha a^{\dagger}-\alpha^{\ast}a}$ and $
D_{b}(\beta)=e^{\beta b^{\dagger}-\beta^{\ast}b}$.

For the input state $\rho_{\rm{in}}$, the output state is $\rho_{\rm{out}}=U(\phi)\rho_{\rm{in}}U^{\dagger}(\phi)$,
where $U(\phi)=e^{-i\phi J_{y}}$ is the unitary evolution operator of the MZI.
The Wigner function of the output state is
\begin{eqnarray}
&&W_{\rm{out}}(\alpha,\beta)\nonumber\\
&=&\frac{4}{\pi^{2}}\textrm{Tr}[
\rho_{\rm{out}}D_{b}(\beta)D_{a}(\alpha)(-1)^{a^{\dagger}a+b^{\dagger}b}D_{a}^{\dagger}(\alpha)D_{b}^{\dagger}(\beta)]\notag\\
&=&\frac{4}{\pi^{2}}\textrm{Tr}[\rho_{\rm{in}}\Lambda(\phi,\alpha,\beta)(-1)^{a^{\dagger}a+b^{\dagger}b}\Lambda^{\dagger}(\phi,\alpha,\beta)]\label{wignfunout}
\end{eqnarray}
with
\begin{eqnarray}
\Lambda(\phi,\alpha,\beta)=U^{\dagger}(\phi)D_{b}(\beta)D_{a}(\alpha)U(\phi).
\end{eqnarray}
In Eq.~(\ref{wignfunout}), we have used the commutation relation $[a^{\dagger}a+b^{\dagger}b,J_{y}]=0$.

In terms of the relations
\begin{subequations}
\begin{align}
U^{\dagger}(\phi)aU(\phi)&=a\cos(\phi/2)-b\sin(\phi/2),\\
U^{\dagger}(\phi)bU(\phi)&=a\sin(\phi/2)+b\cos(\phi/2),
\end{align}
\end{subequations}
we obtain
\begin{eqnarray}
\Lambda(\phi,\alpha,\beta)=D_{a}(\tilde{\alpha})D_{b}(\tilde{\beta}),
\end{eqnarray}
where we introduce
\begin{subequations}
\begin{eqnarray}
\tilde{\alpha}&=&\alpha\cos(\phi/2)+\beta\sin(\phi/2),\\
\tilde{\beta}&=&-\alpha\sin(\phi/2)+\beta\cos(\phi/2).
\end{eqnarray}
\end{subequations}
Therefore, the Wigner function of the output state can be expressed as
$W_{\rm{out}}(\alpha,\beta)=W_{\rm{in}}(\tilde{\alpha},\tilde{\beta})$.

\end{document}